\documentclass[12pt]{article}

\makeatletter
\@addtoreset{equation}{section}
\makeatother

\def\ben{\begin{equation}}
\def\een{\end{equation}}

  \let\n=\nu

\let\C=\Chi

\def\nn{\nonumber} \def\bd{\begin{document}} \def\ed{\end{document}}
\def\ds{\documentstyle} \let\fr=\frac \let\bl=\bigl \let\br=\bigr
\let\Br=\Bigr \let\Bl=\Bigl
\let\bm=\bibitem
\let\na=\nabla
\let\pa=\partial \let\ov=\overline
\newcommand{\be}{\begin{equation}}
\newcommand{\ee}{\end{equation}}
\def\ba{\begin{array}}
\def\ea{\end{array}}
\def\ft#1#2{{\textstyle{{\scriptstyle #1}\over {\scriptstyle #2}}}}
\def\fft#1#2{{#1 \over #2}}
\def\del{\partial}
\def\vp{\varphi}
\def\sst#1{{\scriptscriptstyle #1}}
\def\oneone{\rlap 1\mkern4mu{\rm l}}
\def\td{\tilde}
\def\wtd{\widetilde}
\def\ie{\rm i.e.\ }
\def\dalemb#1#2{{\vbox{\hrule height .#2pt
        \hbox{\vrule width.#2pt height#1pt \kern#1pt
                \vrule width.#2pt}
        \hrule height.#2pt}}}
\def\square{\mathord{\dalemb{6.8}{7}\hbox{\hskip1pt}}}
\newcommand{\ho}[1]{$\, ^{#1}$}
\newcommand{\hoch}[1]{$\, ^{#1}$}
\newcommand{\bea}{\begin{eqnarray}}
\newcommand{\eea}{\end{eqnarray}}
\newcommand{\ra}{\rightarrow}
\newcommand{\lra}{\longrightarrow}
\newcommand{\Lra}{\Leftrightarrow}
\newcommand{\ap}{\alpha^\prime}
\newcommand{\bp}{\tilde \beta^\prime}
\newcommand{\tr}{{\rm tr} }
\newcommand{\Tr}{{\rm Tr} }
\def\0{{\sst{(0)}}}
\def\1{{\sst{(1)}}}
\def\2{{\sst{(2)}}}
\def\3{{\sst{(3)}}}
\def\4{{\sst{(4)}}}
\def\5{{\sst{(5)}}}
\def\6{{\sst{(6)}}}
\def\7{{\sst{(7)}}}
\def\8{{\sst{(8)}}}
\def\n{{\sst{(n)}}}
\def\cA{{{\cal A}}}
\def\cB{{{\cal B}}}
\def\cF{{{\cal F}}}
\def\tV{\widetilde V}
\def\tW{\widetilde W}
\def\tH{\widetilde H}
\def\tE{\widetilde E}
\def\tF{\widetilde F}
\def\tA{\widetilde A}
\def\im{{{\rm i}}}
\def\tY{{{\wtd Y}}}
\def\ep{{\epsilon}}
\def\vep{{\varepsilon}}
\def\R{\rlap{\rm I}\mkern3mu{\rm R}}
\def\bD{{{\bar D}}}

\def\R{\rlap{\rm I}\mkern3mu{\rm R}}
\def\bD{{{\bar D}}}
\def\R{{{\Bbb R}}}
\def\C{{{\Bbb C}}}
\def\H{{{\Bbb H}}}
\def\CP{{{\Bbb C}{\Bbb P}}}
\def\RP{{{\Bbb R}{\Bbb P}}}
\def\Z{{{\Bbb Z}}}
\def\bA{{{\Bbb A}}}
\def\bB{{{\Bbb B}}}
\def\bC{{{\Bbb C}}}
\def\bD{{{\Bbb D}}}
\def\bE{{{\Bbb E}}}
\def\bZ{{{\Bbb Z}}}
\def\Re{{{\frak{Re}}}}
\def\Im{{{\frak{Im}}}}
\def\cosec{{\,\hbox{cosec}\,}}
\def\Gm{{\Gamma_{\!\! -}}}
\def\Gp{{\Gamma_{\!\! +}}}
\def\stan{{standard }}
\def\nonstan{{supernumerary }}

\newcommand{\tamphys}{\it Center for Theoretical Physics,
Texas A\&M University, College Station, TX 77843}

\newcommand{\upenn}{\it Department of Physics and Astronomy,\\ University
of Pennsylvania, Philadelphia, PA 19104}

\newcommand{\brussels}{\it Physique Th\'eorique et Math\'ematique,
Universit\'e Libre de Bruxelles,\\ Campus Plaine C.P. 231, B-1050
Bruxelles, Belgium}

\newcommand{\auth}{J. Kerimo\hoch{\dagger1}, James T. Liu\hoch{\ddagger2}, 
H. L\"u\hoch{\dagger1} and C.N. Pope\hoch{\dagger1}}

\begin{document}
\begin{flushright}

MIFP-04-02\ \ \ MCTP-04-09\\
{\bf  hep-th/0402223}\\
February\  2004
\end{flushright}

\vspace{10pt}

\begin{center}

{\large {\bf Supergravities with Minkowski $\times$  Sphere Vacua}}

\vspace{20pt}
\auth

\vspace{20pt} {\hoch{\dagger} \it George P. and Cynthia W. Mitchell
Institute for Fundamental Physics,\\ Texas A\& M University,
College Station, TX 77843-4242, USA}

\vspace{10pt} {\hoch{\ddagger}
\it Michigan Center for Theoretical Physics\\
University of Michigan, Ann Arbor, MI 48109-1120}

\vspace{40pt}

\underline{ABSTRACT}\\
\end{center}

Recently the authors have introduced a new gauged supergravity theory
with a positive definite potential in $D=6$, obtained through a
generalised Kaluza-Klein reduction from $D=7$.  Of particular interest
is the fact that this theory admits certain Minkowski$\times$Sphere
vacua.  In this paper we extend the previous results by constructing
gauged supergravities with positive definitive potentials in diverse
dimensions, together with their vacuum solutions.  In addition, we
prove the supersymmetry of the generalised reduction ansatz. We obtain
a supersymmetric solution with no form-field fluxes in the new gauged
theory in $D=9$.  This solution may be lifted to $D=10$, where it
acquires an interpretation as a time-dependent supersymmetric
cosmological solution supported purely by the dilaton.  A further
uplift to $D=11$ yields a solution describing a pp-wave.

{\vfill\leftline{}\vfill \vskip 10pt \footnoterule {\footnotesize
\hoch{1} Research supported in part by DOE grant
DE-FG03-95ER40917.}\vskip -1pt
\footnotesize \hoch{2} Research supported in part by DOE grant
DE-FG02-95ER40899.}

\pagebreak
\setcounter{page}{1}

\tableofcontents
\addtocontents{toc}{\protect\setcounter{tocdepth}{2}}
\newpage

\section{Introduction}

    Recent interest in both de Sitter and anti-de Sitter vacua has led
to a renewed study of gauged supergravities, where the gauging of some
$R$-symmetry naturally leads to a non-trivial potential.  Well-known
examples include the gauged supergravities in four, five and seven
dimensions that admit maximally supersymmetric anti-de Sitter vacua.
In addition, there are also gauged supergravities with run-away
potentials.  Although such theories do not admit maximally supersymmetric
vacua, they typically allow domain-wall solutions where scalar
gradient energy is balanced against the scalar potential.  What has
not been achieved, however, is the construction of conventional gauged
supergravities admitting de Sitter vacua.  Of course this is not
particularly surprising, since de Sitter spacetime is incompatible with
conventional supersymmetry.

    Supergravities with positive-definite (albeit run-away) potentials
do nevertheless exist.  A particularly interesting example is the
Salam-Sezgin model, which is a gauged ${\cal N}=(1,0)$ supergravity in
$D=6$ coupled to a tensor and an abelian vector multiplet
\cite{szmodel}.  This model has a (Minkowski)$_4\times S^2$ vacuum, in
which the vector has a non-trivial flux on the $2$-sphere.  This
monopole flux, combined with the single-exponential potential $V\sim
\exp(-\varphi/\sqrt{2})$, is responsible for a ``self-tuning'' of the
vacuum, in which the positive energy density is confined to the
$2$-sphere, thereby ensuring a vanishing $4$-dimensional cosmological
constant and correspondingly a (Minkowski)$_4$ vacuum.  The
self-tuning feature of this model has attracted much attention,
especially as a means of protecting the cosmological constant from
large corrections even after supersymmetry breaking
\cite{abpq1,abpq2}.  It was shown in \cite{cgp} that the Salam-Sezgin
theory arises from a consistent reduction of ten-dimensional
supergravity on a circle times a hyperbolic 3-space.  It was also
shown, in \cite{gp}, that the Salam-Sezgin model can be consistently
reduced on $S^2$ to give rise to ${\cal N}=1$, $D=4$ supergravity
coupled to an $SU(2)$ vector multiplet and a scalar multiplet.

   The interesting features of the Salam-Sezgin model have led us to
search for other possible supergravity theories with positive-definite
potentials.  This search was guided by the realization of \cite{lav}
that a generalised Kaluza-Klein reduction which gauges a combination
of a homogeneous global scaling symmetry together with a Cremmer-Julia
type global symmetry yields a consistent reduction with just such a
positive-definite potential.  In particular, this generalised reduction
was used to construct a variant ${\cal N}=(1,1)$ supergravity in $D=6$
admitting both (Minkowski)$_4\times S^2$ and (Minkowski)$_3\times S^3$ 
vacua \cite{kerimo1,kerimo2}.  This construction is based on the generalised
reduction of minimal $D=7$ supergravity, where a would-be vector multiplet
may be truncated out by a judicious choice of the gauging parameters.
In this manner, the reduction takes one from a pure $(d+1)$-dimensional
supergravity without a potential to a pure $d$-dimensional supergravity
with a (positive-definite) single-exponential potential.  Generalised
Kaluza-Klein reduction {\it via} the gauging of the Cremmer-Julia global
symmetries were considered in \cite{brgpt,clpst,mo,berg}

Although the work of \cite{kerimo1,kerimo2} focused on the reduction
from seven to six dimensions, the generalised Kaluza-Klein procedure
may be carried out in arbitrary dimensions.  In general, the various
supergravities in diverse dimensions are quite distinct (especially in
their fermionic sectors). However it is noteworthy that the bosonic
sector of the half-maximal (16 supercharge) supergravities in $D\le10$ is
universal, with field content
\begin{equation}
(g_{\mu\nu}, B_{\mu\nu}, \phi, A_\mu^a)
\end{equation}
($a=1,2,\ldots,10-D$).  This is of course the bosonic content of the
heterotic string (or the NS-NS sector of the Type-II string) compactified
on a $(10-D)$-dimensional torus, with vector multiplets truncated out.
Owing to this universality of the field content, we may perform a generalised
Kaluza-Klein reduction on the half-maximal supergravities in arbitrary
dimensions, and in this manner obtain the full class of (16
supercharge) variant supergravities generalising the results of
\cite{kerimo1,kerimo2}.

   The resulting $d$-dimensional variant supergravities admit both
(Minkowski)$_{d-3}\times S^3$ and also, in certain cases,
(Minkowski)$_{d-2}\times S^2$, vacua.  Furthermore, we are able to
construct a new time-dependent supersymmetric solution (or
``cosmological solution'') in $D=9$ with no form-field fluxes.  This
solution lifts to a purely dilaton driven cosmology in $D=10$, and a
pp-wave in $D=11$.

\section{Generalised reduction}\label{genredsec}

    We begin with the generalised Kaluza-Klein reduction of the bosonic
sector of half-maximal supergravities in arbitrary dimensions $D\le10$.
In this section, all fields and their equations of motion pertain to
the Einstein frame.  The string-frame picture will be examined in
section 3.

    As indicated above, the bosonic field content of pure supergravity
with 16 supercharges consists of the graviton $\hat g_{\mu\nu}$,
antisymmetric tensor $\hat B_{\mu\nu}$ and dilaton $\hat\phi$, along
with $(10-D)$ $1$-form potentials $\hat A_\mu^a$.  The Lagrangian for
the bosonic sector can be written as
\be
\hat{\cal L}=\hat R\hat\ast\oneone-\ft12\hat\ast d\hat\phi\wedge
d\hat\phi-\ft12 e^{\hat a\hat\phi}\hat\ast\hat H_\3\wedge\hat H_\3
-\ft12 e^{\fft12 \hat a\hat \phi}\hat\ast\hat F_\2^a\wedge\hat F^a_\2
\,,\label{dlag}
\ee
where $\hat F_\2^a=d\hat A_\1^a$, $\hat H_\3 = d\hat B_\2 -\ft12 \hat
F_\2^a\wedge \hat A_\1^a$, and $a=1,2,\ldots, (10-D)$.  The constant
$\hat a$ is given by
\be
\hat a^2= \fft{8}{D-2}\,.
\ee
The equations of motion following from (\ref{dlag}) are
\bea
\hat R_{\sst{MN}}&=&\ft12\del_{\sst M}\hat\phi\,\del_{\sst N}\hat\phi
+\ft14e^{\hat a\hat\phi}\Bigl(\hat H_{\sst {MPQ}}\, 
\hat H_{\sst N}^{\;\,\sst{PQ}} -\frac2{3(D-2)}\hat H^2_\3\,
\hat g_{\sst{MN}}\Bigr)\nn\\
&& +\ft12e^{\fft12\hat a\hat\phi}\Bigl(\hat F^a_{\sst {MP}}\,
\hat F^{a\;\sst P}_{\sst N}-\frac1{2(D-2)}(\hat F^{a}_\2)^2\,
\hat g_{\sst{MN}}\Bigr)\,,\nn\\
d(e^{\hat a\hat\phi}\,\hat\ast\hat H_\3)&=&0\,,\nn\\
d(e^{\fft12 \hat a\hat\phi}\,\hat\ast\hat F^a_\2)&=&
{(-1)}^{D+1}e^{\hat a\hat\phi}\,\hat\ast\hat H_\3
\wedge\hat F^a_\2\,,\nn\\
\widehat{\square}\hat\phi&=&\frac{\hat a}{12}\,
e^{\hat a\hat\phi}\hat H_\3^2
+\frac{\hat a}8\,e^{\fft12\hat a\hat\phi}\,
(\hat F^{a}_\2)^2\,.\label{dgeneom}
\eea

The key observation behind the generalised reduction of
ref. \cite{lav} is that the equations of motion are
invariant under the two global symmetries 
\bea
\hat \phi \rightarrow \hat \phi + \frac1{\hat a}\, \lambda_1\,,\qquad
d\hat s^2 \rightarrow e^{2\lambda_2}\, d\hat s^2\,,\nn\\
\hat B_\2\rightarrow e^{-2\lambda_1+2\lambda_2}\, \hat B_\2
\,,\qquad
\hat A_\1^a\rightarrow e^{-\lambda_1 + \lambda_2}\, \hat
A_\1^a\,.
\eea
The constant $\lambda_1$ parameterises a global symmetry of
the Lagrangian, while the scaling transformation parameterised by
the constant $\lambda_2$ is a symmetry only at the level of the
equations of motion, since the Lagrangian scales homogeneously as 
$\sqrt{-\hat g}(\hat R+\cdots)\longrightarrow e^{(D-2)\, \lambda_2}\,
\sqrt{-\hat g}\, (\hat R+\cdots)$.

   Following \cite{lav}, we now reduce from $D$ dimensions to
$d=(D-1)$, while simultaneously gauging the above two
global symmetries.  The $D$-dimensional pure supergravity multiplet
then reduces to $d$-dimensional supergravity coupled to a single
vector multiplet.  This is achieved by making the
generalised reduction ansatz
\bea
d\hat s^2 &=& e^{2m_2z}\, \Big(e^{2\alpha \varphi}\, ds^2 +
e^{2\beta \varphi}\, (dz + \cA_\1)^2\Big)\,,\nn\\
\hat B_\2 &=& e^{2(m_2-m_1)z}\, \Big(B_\2 + B_\1\wedge dz\Big)\,,\nn\\
\hat A_\1^a &=& e^{(m_2-m_1)z}\, \Big(A_\1^a + \chi^a\, dz\Big)\,,\nn\\
\hat \phi &=& \phi + \frac4{\hat a}\, m_1 z\,, \label{genredans}
\eea
where
\be
{\alpha}^2=\frac{1}{2(d-1)(d-2)}\,,\qquad \beta=-(d-2)\alpha\,. \label{diag}
\ee
The standard Kaluza-Klein ansatz for an ungauged $S^1$ reduction
would correspond to setting $m_1=m_2=0$.

   In general, for unequal mass parameters $m_1$ and $m_2$, the
lower-dimensional equations of motion are rather complicated.
However, a significant simplification occurs if $m_1=m_2$.  In
this case, various exponential factors drop out from
(\ref{genredans}), and one can consistently truncate out
the vector multiplet, owing to conspiracies between the
fields.  In this manner, one can obtain variant gauged supergravities
with positive-definite scalar potentials and with half-maximal
supersymmetry in $d\le9$ dimensions.

   Before writing out the complete reduction of the bosonic equations of
motion, we first collect some intermediate results.  The
reduction of the potentials in (\ref{genredans}) yields a
corresponding reduction on the field strengths:
\bea
\hat H_\3&=&e^{2(m_2-m_1)z}(H_\3+H_\2\wedge(dz+\cA_\1))\,,\nn\\
\hat F^a_\2&=&e^{(m_2-m_1)z}(F^a_\2+L^a_\1\wedge(dz+\cA_\1))\,,
\label{fieldstrengths}
\eea
where the lower dimensional fields are defined by
\bea
H_\3&=&dB_\2-\ft{1}{2}F^a_\2{\wedge}A^a_\1-dB_\1\wedge
\cA_\1-2(m_2-m_1)B_\2\wedge\cA_\1+\ft12\chi^a F^a_\2
\wedge\cA_\1\,,\nn\\
G_\2&=&dB_\1-\ft{1}{2}\chi^a\,F^a_\2+\ft12L^a_\1
\wedge A^a_\1-\ft12\chi^a L^a_\1\wedge\cA_\1+2(m_2-m_1)B_\2\,,\nn\\
F^a_\2&=&dA^a_\1-d\chi^a\wedge\cA_\1+(m_2-m_1)A^a_\1\wedge
\cA_\1\,,\nn\\
L^a_\1 &=& d\chi^a-(m_2-m_1)A^a_\1
\,.\label{dfe}
\eea
The Kaluza-Klein potential $\cA_\1$ has the standard field
strength $\cF_\2=d\cA_\1$.  It is evident at this stage that the
vector fields $A^a_\1$ and the tensor field $B_\2$ acquire masses
proportional to $| m_2-m_1|$, in the process eating the axions
$\chi^a$ and the vector $B_\1$ respectively.

\subsection{Untruncated $d$-dimensional equations}

   We are now able to write down the
full bosonic equations of motion for the variant $d$-dimensional gauged
supergravity. The bosonic field content is
\begin{equation}
(g_{\mu\nu}, B_{\mu\nu},\varphi,A_\mu^a,\cA_\mu)\quad\hbox{and}\quad
(B_\mu,\chi^a,\phi)\,,
\label{eq:multiplets}
\end{equation}
corresponding to half-maximal supergravity coupled to a single vector
multiplet.  This representation is schematic in the sense that the
scalars $\phi$ and $\varphi$ as well as the $1$-form potentials $B_\1$
and $\cA_\1$ must necessarily be taken as appropriate linear
combinations in the actual multiplets.

   We find that the equations of motion for the form fields are given by
\bea
\nabla^\sigma(e^{\hat a\phi-4\alpha\varphi}H_{\mu\nu\sigma})&=&
(2m_1+(d-3)\,m_2)\,{\Big(}e^{\hat a\phi-4\alpha\varphi}
H_{\mu\nu\sigma}\cA^{\,\sigma}-e^{\hat a\phi+2(d-3)\alpha\varphi}
G_{\mu\nu}{\Big)}\,,\nn\\
 \nabla^\nu(e^{\hat a\phi+2(d-3)\alpha\varphi}G_{\mu\nu})&=&\ft12
e^{\hat a\phi-4\alpha\varphi}H_{\mu\nu\sigma}\cF^{\nu\sigma}\nn\\
&& +\,(2m_1+(d-3)\,m_2)e^{\hat a\phi+2(d-3)\alpha\varphi}
G_{\mu\nu}\,\cA^\nu\,,\nn\\
\nabla^\nu(e^{\fft12\hat a\phi-2\alpha\varphi}F^a_{\mu\nu})&=&
\ft12e^{\hat a\phi-4\alpha\varphi}H_{\mu\nu\sigma}
F^{a\nu\sigma}+e^{\hat a\phi+2(d-3)\alpha\varphi}G_{\mu\nu}L^{a\nu}\nn\\
&& +\,(m_1+(d-2)\,m_2)
{\Big(}e^{\fft12\hat a\phi-2\alpha\varphi}F^a_{\mu\nu}
\cA^\nu-e^{\fft12\hat a\phi+2(d-2)\alpha\varphi}L^a_\mu{\Big)}
\,,\nn\\
\nabla^\mu(e^{\fft12\hat a\phi+2(d-2)\alpha\varphi}L^a_\mu)
 &=& -\,\ft12e^{\hat a\phi+2(d-3)\alpha\varphi}G_{\mu\nu}\,
F^{a\mu\nu}+\ft12e^{\fft12\hat a\phi-2\alpha\varphi}
F^a_{\mu\nu}\cF^{\mu\nu}\nn\\
&& +\,(m_1+(d-2)\,m_2)e^{\fft12\hat a\phi+2(d-2)\alpha\varphi}
L^a_\mu\cA^\mu\,,\nn\\
\nabla^\nu(e^{-2(d-1)\alpha\varphi}\cF_{\mu\nu}) &=&
\ft12e^{\hat a\phi-4\alpha\varphi}H_{\mu\nu\sigma}G^{\,\nu\sigma}
-e^{\fft12\hat a\phi-2\alpha\varphi}F^a_{\mu\nu}L^{a\nu}\nn\\
&& +\,\frac4{\hat a}\,m_1(\del_{\mu}\phi-
\frac4{\hat a}\,m_1\cA_{\mu}) -2m_2(d-1)\,
(\beta\del_{\mu}\varphi-m_2\cA_{\mu})\nn\\
&&+\,m_2(d-1)e^{-2(d-1)\alpha\varphi}\,
\cF_{\mu\nu}\cA^\nu\,.
\eea

   The two scalar fields, $\phi$ and $\varphi$ satisfy similar $m_1$ and
$m_2$ dependent equations of motion.  The scalar coming from the 
metric satisfies the equation 
\bea
-\beta\,\square\varphi&=&-\,\frac{e^{\hat a\phi-4\alpha\varphi}}{6(d-1)}
H^{2}_\3-\,\frac{e^{\fft12\hat a\phi-2\alpha\varphi}}{4(d-1)}
(F^{a}_\2)^2+\,\frac{d-3}{4(d-1)}e^{\hat a\phi+2(d-3)\alpha\varphi}\,
G^{2}_\2\nn\\
&&+\,\frac{d-2}{2(d-1)}\,e^{\fft12\hat a\phi+2(d-2)\alpha\varphi}
(L^a_\1)^2-\ft14e^{-2(d-1)\alpha\varphi}\cF_\2^2\label{wave1}\\
&&-\,m_2\beta(d-1)\cA^\mu\del_{\mu}\varphi-
m_2\nabla_\mu\cA^\mu+m_2^2(d-1){\cal A}_\1^2
+\frac8{\hat a^2}\,m_1^2e^{2(d-1)\alpha\varphi}\,,\nn
\eea
while the $D$-dimensional dilaton equation reduces to
\bea
\square\phi&=&\frac{\hat a}{12}\,e^{\hat a\phi-4\alpha\varphi}
H^{2}_\3+\frac{\hat a}4\,e^{\hat a\phi+2(d-3)\alpha\varphi}
G^{2}_\2+\,\frac{\hat a}8\,e^{\fft12\hat a\phi-2\alpha\varphi}
(F^{a}_\2)^2\nn\\
&&+\frac{\hat a}4\,e^{\fft12\hat a\phi+2(d-2)\alpha\varphi}
(L^{a}_\1)^2+\,m_2(d-1)\cA^\mu\del_\mu\phi+\frac4{\hat a}\,m_1
\nabla_\mu\cA^\mu\nn\\
&&-\frac{4(d-1)}{\hat a}\,m_1m_2\,
(\cA_\1^2+e^{2(d-1)\alpha\varphi})\,.\label{wave2}
\eea

   The $d$-dimensional Einstein equation takes the form
\bea
&&R_{\mu\nu}-\ft12Rg_{\mu\nu}=\ft12 (\del_\mu\varphi\,
\del_\nu \varphi - \ft12 (\del\varphi)^2\, g_{\mu\nu})+
\ft12 (\del_\mu\phi\, \del_\nu \phi - \ft12 
(\del\phi)^2\, g_{\mu\nu})\nn\\
&&+\,\ft12e^{-2(d-1)\alpha\varphi}\,
(\cF_{\mu\sigma}\cF_\nu^{\;\,\sigma}-
\ft14g_{\mu\nu}\cF_\2^2)+\ft14
e^{\hat a\phi-4\alpha\varphi}(H_{\mu\rho\sigma}\,
H_{\nu}^{\;\,\rho\sigma}-\ft16g_{\mu\nu}H^{2}_\3)\nn\\
&&+\,\ft12e^{\fft12\hat a\phi-2\alpha\varphi}
(F^a_{\mu\sigma}\,F^{a\,\sigma}_\nu-\ft14g_{\mu\nu}
(F^a_\2)^2)+\ft12e^{\hat a\phi+2(d-3)\alpha\varphi}
(G_{\mu\sigma}G_\nu^{\;\,\sigma}-\ft14
g_{\mu\nu}G^{2}_\2)\nn\\
&&+\,\ft12e^{\fft12\hat a\phi+2(d-2)\alpha\varphi}\,
(L^a_\mu L^a_\nu-\ft12g_{\mu\nu}\,(L^a_\1)^2)\nn\\
&&-\alpha m_2(d-1)(\cA^\sigma\del_\sigma\varphi\,
g_{\mu\nu}-\cA_\mu\del_\nu\varphi-
\cA_\nu\del_\mu\varphi)\nn\\
&&+\,\frac2{\hat a}\,m_1(\cA^\sigma\del_\sigma\phi \,
g_{\mu\nu}-\cA_\mu\del_\nu\phi-\cA_\nu\del_\mu\phi)+
{\Big(}\frac8{\hat a^2}\,m_1^2-(d-1)m_2^2{\Big)}
\cA_\mu\cA_\nu\nn\\
&&-\,\ft12m_2(d-1)(\nabla_\mu\cA_\nu+
\nabla_\nu\cA_\mu-2\nabla_\sigma\cA^\sigma\,
g_{\mu\nu})\nn\\
&&-\,{\Big(}\frac{4m_1^2}{\hat a^2}+\ft12m_2^2(d-1)(d-2){\Big)}
\,(\cA_\1^2+e^{2(d-1)\alpha\varphi})g_{\mu\nu}\,.\label{ei}
\eea
Note that the last term is associated with a positive-definite 
scalar potential.

\subsection{Truncated $d$-dimensional equations}

   The scalars $\phi$ and $\varphi$ may be disentangled between the
supergravity and vector multiplets of (\ref{eq:multiplets}) by
performing a rotation to $\phi_1$ (supergravity) and $\phi_2$ (vector)
given by
\be
\hat a\phi - 4\alpha\varphi =a\phi_1\,,\qquad
4\alpha\phi + \hat a\varphi =a\phi_2\,,
\ee
where $a=\sqrt{8/(D-3)}$.  When $m_1=m_2$, the vector multiplet
may be further truncated away.  This is done by setting
\be
B_\1=\cA_\1\equiv \ft1{\sqrt2} A_\1\,,\qquad
\phi_2=0\,,\qquad L_\1^a=0\,.
\ee

The equations of motion for the pure supergravity fields are then
given by
\bea
&&\nabla^\rho\Big(e^{a\phi}H_{\mu\nu\rho}\Big) =
\frac{d-1}{\sqrt2}\,m \Big(e^{a\phi}\,
H_{\mu\nu\rho}\, A^\rho - e^{\fft12a\phi}F_{\mu\nu}\Big)
\,,\nn\\
&&\nabla^\nu\Big(e^{\fft12a\phi}F_{\mu\nu}\Big) =
\ft12 e^{a\phi}H_{\mu\nu\rho}F^{\nu\rho} +
\frac{d-1}{\sqrt2}\,m e^{\fft12a\phi}F_{\mu\nu}\,A^\nu\,,\nn\\
&&\nabla^\nu\Big(e^{\fft12a\phi}F^a_{\mu\nu}\Big)=
\ft12e^{a\phi}H_{\mu\nu\rho}\,F^{a\,\nu\rho} +
\frac{d-1}{\sqrt2}\,m e^{\fft12a\phi}F^a_{\mu\nu}\,
A^\nu\,,\nn\\
&&\square\phi = \fft{e^{a\phi}}{3\sqrt{2(d-2)}} H_\3^2 +
\frac{e^{\fft12a\phi}}{2\sqrt{2(d-2)}} (F_\2^2 + (F_\2^a)^2)
+\frac{d-1}{\sqrt2}\,mA^\mu\,\del_\mu \phi\nn\\
&&\qquad\qquad + \frac{d-1}{\sqrt{d-2}}\, m \nabla_\mu\,A^\mu -
\frac{\sqrt2\, (d-1)^2}{\sqrt{d-2}}\, m^2 (\ft12 A_\1^2+
e^{-\fft12a\phi})\,,\nn\\
&&R_{\mu\nu}=\ft12\del_\mu\phi\del_\nu\phi+\ft14e^{a\phi}
(H_{\mu\rho\sigma}H_{\nu}^{\;\,\rho\sigma}
-\frac2{3(d-2)}H^{2}_\3g_{\mu\nu})\nn\\
&&+\ft12e^{\fft12a\phi}(F_{\mu\rho}F_\nu^{\;\,\rho}
-\frac1{2(d-2)}F^{2}_\2g_{\mu\nu})
+\ft12e^{\fft12a\phi}(F^a_{\mu\rho}F_\nu^{a\;\rho}
-\frac1{2(d-2)}(F^a_\2)^2g_{\mu\nu})\nn\\
&&-\frac{m(d-1)}{2\sqrt{d-2}}(A_\mu\del_\nu\phi+A_\nu\del_\mu\phi)
-\frac{m(d-1)}{2\sqrt2}(\nabla_\mu A_\nu+\nabla_\nu A_\mu+\frac2{d-2}
\nabla_\rho A^\rho g_{\mu\nu})\nn\\
&&+\frac{m^2(d-1)^2}{2(d-2)}(A_\1^2+2e^{-\fft12a\phi})g_{\mu\nu}\,,
\eea
where we have  rewritten $\phi_1$ as $\phi$.  
It may be seen that this set of equations cannot be obtained
from a Lagrangian in terms of the physical fields.  This is not
altogether surprising, since they were derived in a generalised
reduction that gauged a symmetry of the equations of motion which was
not a symmetry of the Lagrangian.

   By examining the linearised equations of motion, it can be seen that
$A_\1$ is a massless gauge potential.  This gauge field can in fact
be consistently set to zero.  In this case, the remaining equations of
motion can then be obtained from the Lagrangian
\be
e^{-1}{\cal L} = R - \ft12(\del\phi)^2 - \ft1{12}e^{a\phi}
H_\3^2 - \ft14 e^{\fft12a\phi} (F_\2^a)^2 - (d-1)^2\, m^2
e^{-\fft12a\phi}\,,\label{ddlag1}
\ee
where $e=\sqrt{-g}$.  Thus we see once again that the scalar potential
is positive definite.

\section{String frame and $\sigma$-model action}

   For many purposes it is advantageous to perform the Weyl rescaling of the
metric that transforms from the Einstein frame that we used in the previous
section to the string frame.  One reason is because the half-maximal 
supergravities that we are considering have a direct relation to
the heterotic string, or the NS-NS sector of the
Type-II string.  Another reason is that many of the formulae become 
considerably simpler when expressed in the string frame.  
We shall consider only the case $m_1=m_2=m$.

   Consistent string propagation demands world-sheet conformal
invariance, and hence the vanishing of the beta functions for the
background spacetime fields.  In this manner one obtains
supergravity equations of motion which arise naturally in the
string frame.  The corresponding equations may be derived from the
string-frame Lagrangian
\be
\hat e^{-1}\hat {\cal L}=e^{-2\hat\Phi}(\hat R+4(\del\hat\Phi)^2
-\ft1{12}\hat H^{2}_\3-\ft14(\hat F^a_\2)^2)\,,\label{stringlag}
\ee
taken here to have been compactified on a $(10-D)$-dimensional torus
(with the additional truncation of $(10-D)$ vector multiplets).  It is to be
understood that all fields in this section are labelled with a
suppressed tilde $(\td g_{\mu\nu}\,,\,\wtd H_\3,\mbox{\it etc.})$
unless otherwise indicated, to distinguish them from the Einstein
frame fields.  The complete transformation between the two frames in
dimensions $D\le10$ is given in appendix C.

   The equations of motion following from the Lagrangian (\ref{stringlag}) 
are
\bea
\hat R_{\sst{MN}}&=&-2\widehat\nabla_{\sst M}
\widehat\nabla_{\sst N}\hat\Phi
+\ft14\hat H_{\sst{MPQ}}\hat H_{\sst N}^{\;\,\sst{PQ}}
+\ft12\hat F^a_{\sst{MP}}\hat F_{\sst N}^{a\;\sst P}\,,\nn\\
d(e^{-2\hat\Phi}\hat\ast\hat H_\3)&=&0\,,\nn\\
d(e^{-2\hat\Phi}\hat\ast\hat F^a_\2)&=&(-1)^{D+1}\, e^{-2\hat\Phi}
\hat\ast\hat H_\3\wedge\hat F^a_\2\,,\nn\\
\widehat{\square}\hat\Phi&=&2(\del\hat\Phi)^2
-\ft1{12}\hat H^{2}_\3-\ft18(\hat F^a_\2)^2\,.
\eea
By tracing the Einstein equation and substituting in the dilaton equation,
we may obtain an expression for the Ricci scalar:
\be
\hat R=-4(\del\,\hat\Phi)^2+\ft5{12}\hat H_\3^2
+\ft34(\hat F^a_\2)^2\,.
\ee

   In $D$ dimensions, the Einstein-frame and the string-frame metrics are
 related by
\be
d\hat s_{\rm Ein}^2 = e^{\fft12\hat a\hat\phi}\, d\hat s_{\rm str}^2 =
e^{-\fft12\hat a^2 \hat \Phi}\, d\hat s_{\rm str}^2\,,
\ee
where we have defined $\hat \Phi = -\hat \phi/\hat a$ and $\hat\phi$
is the Einstein-frame dilaton field.  For the case where $m_1=m_2$, the
reduction ansatz (\ref{genredans}) converted to the string frame is
rather simple, namely
\bea
d\hat s_{\rm str}^2 &=& ds_{\rm str}^2 + e^{-\sqrt2\varphi}
(dz + \cA_\1)^2\,,\nn\\
\hat B_\2&=&B_\2 + B_\1\wedge dz\,,\nn\\
\hat \Phi &=& \Phi -\ft1{\sqrt8}\, \varphi - \ft12 (d-1)\,mz\,.
\eea
In other words, the reduction is exactly the same as a standard
Kaluza-Klein reduction, except for a linear $z$-dependence in the
dilaton $\hat\Phi$.

    It follows that the $\sigma$-model action for this generalised
circle reduction is given by
\bea
I&=&\fft1{4\pi\,\alpha'}\int d\sigma\,d\tau \Big[
\sqrt{\gamma}\,\gamma^{ij}\,\del_i X^\mu\, \del_j X^\nu\, 
\hat g_{\mu\nu} +\epsilon^{ij}\,\del_i X^\mu\, \del_j X^\nu\, 
\hat B_{\mu\nu}\nn\\
&&\qquad\qquad\qquad 
+\alpha'\hat R\, (\Phi - \ft12(D-2)\,m z)\Big]\,,\nn
\eea
where $ \Phi$, $\hat g_{\mu\nu}$ and $\hat B_{\mu\nu}$
are independent of $z$, and $X^0$ (the circle coordinate) is given by
$X^0 =z$.  However, the $z$ dependence of the string
action implies that $T$-duality is now broken.  This can also be seen
from the low-energy effective action obtained in the previous section,
where the Kaluza-Klein vector $\cA_\1$ and the winding vector $B_\1$
are clearly not on a parallel footing.

\subsection{Untruncated $d$-dimensional string-frame equations}

We give here the complete set of bosonic equations of motion for the
untruncated system, expressed in the string frame.  It will be seen that
these are considerably simpler than the previous expressions that were
obtained in the Einsten frame.

    For the form fields in the string frame we find
\bea
\nabla^\rho(e^{-2\Phi}H_{\mu\nu\rho})&=&m(d-1){\Big(}
e^{-2\Phi}H_{\mu\nu\sigma}\cA^\sigma-e^{-2\Phi+
\sqrt2\varphi}G_{\mu\nu}{\Big)}\,,\nn\\
\nabla^\nu(e^{-2\Phi+\sqrt2\varphi}G_{\mu\nu})&=&\ft12
e^{-2\Phi}H_{\mu\nu\sigma}\cF^{\nu\sigma}+m(d-1)
e^{-2\Phi+\sqrt2\varphi}G_{\mu\nu}\cA^\nu\,,\nn\\
\nabla^\nu(e^{-2\Phi}F^a_{\mu\nu})&=&\ft12e^{-2\Phi}
H_{\mu\nu\sigma} F^{a\,\nu\sigma}+e^{-2\Phi+\sqrt2\varphi}
G_{\mu\nu}L^{a\,\nu}\nn\\
&&+\,m(d-1){\Big(}e^{-2\Phi}F^a_{\mu\nu}\cA^\nu
-e^{-2\Phi+\sqrt2\varphi}L^a_\mu{\Big)}\,,\nn\\
\nabla^\mu(e^{-2\Phi+\sqrt2\varphi}L^a_\mu)&=&\ft12
e^{-2\Phi}F^a_{\mu\nu}\cF^{\mu\nu}-\ft12e^{-2\Phi
+\sqrt2\varphi}G_{\mu\nu}F^{a\,\mu\nu}\nn\\
&&+\,m(d-1)e^{-2\Phi+\sqrt2\varphi}L^a_\mu\cA^\mu\,,\nn\\
\nabla^\nu(e^{-\frac3{\sqrt2}\varphi}\cF_{\mu\nu})&=&
e^{-\frac1{\sqrt2}\varphi}(\ft12H_{\mu\nu\sigma}
G^{\nu\sigma}-F^a_{\mu\nu}L^{a\nu})+2e^{-\frac3{\sqrt2}\varphi}
(\del_\nu\Phi-\ft1{\sqrt8}\del_\nu\varphi)
\cF_\mu^{\;\,\nu}\nn\\
&&+\,m(d-1)(\sqrt2\,e^{-\frac1{\sqrt2}\varphi}\del_\mu\varphi
+e^{-\frac3{\sqrt2}\varphi}\cA^\nu\cF_{\mu\nu})\,.
\eea

   For the scalar fields, we find
\bea
\square\varphi&=&\ft1{2\sqrt2}(e^{\sqrt2\varphi}G_\2^2
-e^{-\sqrt2\varphi}\cF_\2^2)
+\ft1{\sqrt2}e^{\sqrt2\varphi}(L^a_\1)^2
+\,2\del_{\,\mu}\varphi\del^{\,\mu}\Phi+m(d-1)
\cA^\mu\del_\mu\varphi\,,\nn\\
\square\Phi&=&-\,\ft1{12}H_\3^2-\ft18(F^a_\2)^2
-\ft18(e^{\sqrt2\varphi} G_\2^2+
e^{-\sqrt2\varphi} \cF_\2^2)+2(\del\Phi)^2\\
&&+ 2m(d-1) \cA^\mu\,\del_\mu\Phi
-\,\ft12m(d-1)\nabla_\mu\cA^\mu+\ft12m^2(d-1)^2
(\cA_\1^2+e^{\sqrt2\varphi})\,.\nn
\eea

   The Einstein equations in the string frame are given by
\bea
R_{\mu\nu}&=&\ft12\del_\mu\varphi\del_\nu\varphi - 2\nabla_\mu\del_\nu\Phi
+\ft14H_{\mu\rho\sigma}H_\nu^{\;\,\rho\sigma} + \ft12e^{\sqrt2\varphi}
G_{\mu\rho}G_\nu^{\;\,\rho} + \ft12e^{-\sqrt2\varphi}
\cF_{\mu\rho}\cF_\nu^{\;\,\rho}\nn\\
&&+\,\ft12F^a_{\mu\rho}\,F_\nu^{a\;\rho}+\ft12e^{\sqrt2\varphi}L_\mu^aL_\nu^a
- \ft12m(d-1)(\nabla_\mu\cA_\nu + \nabla_\nu\cA_\mu)
\,.
\eea

\subsection{Truncated $d$-dimensional string-frame equations}

   In the string frame, we may again truncate out the vector
multiplet by setting $\varphi=0$, $L_\1^a=0$ and $\cA_\1=B_\1\equiv
A_\1/\sqrt2$.  The equations of motion for the bosonic fields of the
pure supergravity multiplet now become
\bea
\nabla^\sigma H_{\mu\nu\sigma}&=&2H_{\mu\nu\sigma}M^\sigma- \ft1{\sqrt2}\,
m (d-1) F_{\mu\nu}\,,\nn\\
\nabla^\nu F_{\mu\nu}&=&\ft12H_{\mu\nu\sigma}F^{\nu\sigma}
+2F_{\mu\nu}M^\nu\,,\nn\\
\nabla^\nu F^a_{\mu\nu}&=&\ft12H_{\mu\nu\sigma}F^{a\,\nu\sigma}
+2F^a_{\mu\nu}M^\nu\,,\nn\\
\nabla^\mu M _\mu&=&2M_\1^2-\ft1{12}H_\3^2-\ft18(F_\2^2+(F_\2^a)^2)
+\ft12m^2(d-1)^2\,,\nn\\
R_{\mu\nu}&=&-\nabla_\mu M_\nu-\nabla_\nu M_\mu+\ft14H_{\mu\rho\sigma}
H_\nu^{\;\;\rho\sigma}+\ft12(F_{\mu\rho}\,F^{\;\,\rho}_\nu
+F^a_{\mu\rho}\,F^{a\;\rho}_\nu)\,,
\eea
where we have introduced the field 
\be
M_\1=d\Phi+\frac{m(d-1)}{2\sqrt2}A_\1\,.
\ee
It is evident that the massive field $M_\1$ arises because the dilaton
$\Phi$ is eaten by the gauge field $A_\1$\,.

   As in the Einstein frame, these equations cannot be obtained from a
Lagrangian.  However, if we set $A_\1$ to zero, the equations of
motion for the remaining fields can be obtained from a Lagrangian,
given by
\be
e^{-1}{\cal L}=e^{-2\Phi}\,\Big(
R + 4(\del\Phi)^2 - \ft1{12} H_\3^2 - \ft14 (F_\2^a)^2
- (d-1)^2m^2\Big)\,.
\label{eq:sflag}
\ee
Although this truncation is consistent within the bosonic theory, it
cannot be consistent with the full supergravity, as it would be
incompatible with the structure of the supermultiplets.  Nevertheless,
we see from (\ref{eq:sflag}) that in the string frame the scalar
potential becomes a pure positive cosmological constant.

\section{Supersymmetry}

   With the derivation of the bosonic equations of motion both in the
Einstein frame and the string frame completed, we now turn to a
consideration of the supersymmetry transformation rules for these
generalised reductions.  We shall present the results for two cases
in this section.  The first is the variant ten-dimensional massive gauged
supergravity obtained in \cite{lav} by performing a generalised reduction
of eleven-dimensional supergravity.\footnote{Note that this massive
type IIA supergravity \cite{hlw,lav} is not the same as the massive
IIA theory obtained by Romans \cite{romans10}.}  The reduction in this
case involves just the global scaling symmetry of the $D=11$ equations
of motion.  Then, we shall consider the nine-dimensional massive gauged
theory obtained from massless ${\cal N}=1$, $D=10$ supergravity, using
the generalised reduction involving the two global symmetries that
we discussed in section \ref{genredsec}.  Analogous results for the
six-dimensional gauged theory were obtained in detail in \cite{kerimo2}.

\subsection{Massive type IIA supergravity from $D=11$}\label{d10massivesec}

The supersymmetry transformations in $D=11$ are
\bea
&&\delta\hat e_{\sst M}^{\;\;\sst A}=\hat{\bar\epsilon}
\hat\gamma^{\sst A}
\hat\psi_{\sst M}\,, \qquad \delta\hat A_{\sst{MNP}}=3\hat{\bar\epsilon}
\hat\gamma_{[\sst{MN}}\hat\psi_{{\sst P}]}\,,\nn\\
&&\delta\hat\psi_{\sst M}=\widehat\nabla_{\sst M}\hat\epsilon-\ft1{288}
\hat F_{\sst{NPQR}}(\hat\gamma_{\sst M}^{\;\;\sst{NPQR}}-
8\hat\gamma^{\sst{PQR}}
\delta^{\sst N}_{\sst M})\,\hat\epsilon\,,
\eea
where in our conventions 
\be
\{\hat\gamma_{\sst A},\hat\gamma_{\sst B}\}=2\hat\eta_{\sst{AB}}
\ee
and the metric signature is $(- + +\cdots +)$. The equations of motion
of the eleven-dimensional theory are invariant under a scaling symmetry,
which was used in \cite{lav} in a generalised reduction to obtain the
bosonic sector of a massive ten-dimensional supergravity.  Here, we
extend that discussion to include the fermionic sector.  This variant
maximal supersymmetric $D=10$ massive theory \cite{hlw,lav} has also been
considered in \cite{berg}. The corresponding ansatz for the generalised
circle reduction of the fermions is
\bea
\hat\epsilon&=&e^{\fft12m_2z}e^{\fft1{24}\varphi}\,\epsilon\,,\nn\\
\hat\psi_{\sst{11}}&=&\ft{2\sqrt2}3e^{-\fft12m_2z}e^{-\fft1{24}
\varphi}\hat\gamma_{\sst{11}}\lambda\,,\nn\\
\hat\psi_a&=&e^{-\fft12m_2z}e^{-\fft1{24}\varphi}(\psi_a-\ft{\sqrt2}{12}
\gamma_a\lambda)\,.
\eea
Performing the reduction of the fermionic transformation rules, 
we obtain
\bea
\delta\lambda &=& -\,\ft1{2\sqrt2}\gamma^\mu\epsilon
\,\del_\mu\varphi-\ft1{192\sqrt2}e^{-\fft14\varphi}
F_{\mu\nu\sigma\rho}\gamma^{\mu\nu\sigma\rho}\epsilon
+\ft1{24\sqrt2}e^{\fft12\varphi}F_{\mu\nu\sigma}
\gamma^{\mu\nu\sigma}\hat\gamma_{\sst{11}}\epsilon\nn\\
&&-\,\ft3{16\sqrt2}e^{-\fft34\varphi}\cF_{\mu\nu}\gamma^{\mu\nu}
\hat\gamma_{\sst{11}}\epsilon-\ft3{4\sqrt2}m_2(\cA_\mu
\gamma^\mu-e^{\fft34\varphi}\hat\gamma_{\sst{11}})\epsilon\,,\nn\\
\delta\psi_\mu &=& \nabla_\mu\epsilon-\ft1{256}e^{-\fft14\varphi}
F_{\nu\alpha\sigma\rho}{\Big(}\gamma_\mu^{\;\;\nu\alpha\sigma\rho}-\ft{20}3
\delta^{\nu}_{\mu}\gamma^{\alpha\sigma\rho}{\Big)}\epsilon-\ft1{96}e^{\fft12
\varphi}F_{\nu\sigma\rho}{\Big(}\gamma_\mu^{\;\;\nu\sigma\rho}
-9\delta_\mu^\nu\gamma^{\sigma\rho}{\Big)}\hat\gamma_{\sst{11}}\epsilon\nn\\
&&-\ft1{64}e^{-\fft34\varphi}\cF_{\nu\sigma}{\Big(}
\gamma_\mu^{\;\;\nu\sigma}-14\delta^\nu_\mu\gamma^\sigma{\Big)}
\hat\gamma_{\sst{11}}\epsilon-\ft9{16}m_2(\cA_\nu\gamma_\mu\gamma^\nu
-e^{\fft34\varphi}\gamma_\mu\hat\gamma_{\sst{11}})\epsilon\,.\label{d10fermi}
\eea
The supersymmetry transformation rules for the bosons are
\bea
&&\delta e_\mu^{\;\;a}=\bar\epsilon\gamma^a\psi_\mu\,, \qquad
\delta\phi=-\sqrt2\,\bar\epsilon\,\lambda\,,\nn\\
&&\delta {\cal A}_\mu=e^{\fft34\phi}
\bar\epsilon\hat \gamma_{\sst{11}}(\psi_\mu
-\ft{3\sqrt2}4\gamma_\mu\lambda)\,,\nn\\
&&\delta A_{\mu\nu}=e^{-\fft12\phi}\bar\epsilon\hat\gamma_{\sst{11}}
(2\gamma_{[\mu}\psi_{\nu]}+\ft1{\sqrt2}\gamma_{\mu\nu}\lambda)\,,\nn\\
&&\delta A_{\mu\nu\rho}=3e^{\fft14\phi}
\bar\epsilon(\gamma_{[\mu\nu}\psi_{\rho]}
-\ft{\sqrt2}{12}\gamma_{\mu\nu\rho}\lambda)+
3\cA_{[\mu}\delta A_{\nu\rho]}\,.\label{d10bose}
\eea

   As was shown in \cite{lav} this theory admits a de Sitter vacuum
solution, which necessarily breaks all supersymmetry. Note that the ten
dimensional field strengths are those defined in \cite{lav}.

\subsection{Reduction of $D=10, {\cal N}=1$ supersymmetry}

   Since we have obtained the transformation rules for the type IIA
massive gauged supergravity in section \ref{d10massivesec}, it is
convenient to make use of these here in order to establish our conventions
and notation for the transformation rules of the standard massless ${\cal
N}=1$ supergravity in ten dimensions.  These are obtained by setting the
mass parameter $m_2=0$ in (\ref{d10fermi}), and in addition making the
chiral projection that reduces the ${\cal N}=2$ supersymmetry to ${\cal N}=1$:
\be
\hat\gamma_{\sst{11}}\epsilon=\epsilon\,,\qquad
\hat\gamma_{\sst{11}}\psi_a=\psi_a
\qquad \mbox{and} \qquad \hat\gamma_{\sst{11}}\lambda=-\lambda\,.
\label{chiral}
\ee
The chirality condition is consistent with setting to zero both the 3-form
potential and the Kaluza-Klein vector.  This yields the ten-dimensional
${\cal N}=1$ supersymmetry transformation rules
\bea
\delta\hat\lambda&=&-\,\ft1{2\sqrt2}\,\hat\gamma^{\sst M}\hat\epsilon
\,\del_{\sst M}\hat\phi+\ft1{24\sqrt2}\,e^{\fft12\hat\phi}
\,\hat H_{\sst{MNP}}\hat\gamma^{\sst{MNP}}\,\hat\epsilon\,,\nn\\
\delta\hat\psi_{\sst M}&=&\widehat\nabla_{\sst M}\hat\epsilon-
\ft1{96}\,e^{\fft12
\hat\phi}\hat H_{\sst{NPQ}}{\Big(}\hat\gamma_{\sst M}^{\;\;\sst{NPQ}}
-9\,\hat\gamma^{\sst{PQ}}\delta^{\,\sst N}_{\sst M}{\Big)}
\hat\epsilon\,,\nn\\
\delta\hat e_{\sst M}^{\;\;\sst A}&=&\hat{\bar\epsilon}\hat\gamma^{\sst A}
\hat\psi_{\sst M}\,, \qquad \delta\hat\phi=-\sqrt2\,\hat{\bar\epsilon}\,
\hat\lambda\,,\nn\\
\delta\hat B_{\sst{MN}}&=&-e^{-\fft12\hat\phi}\hat{\bar\epsilon}\,
(2\hat\gamma_{[\sst M}
\hat\psi_{\sst N]}+\ft1{\sqrt2}\hat\gamma_{\sst{MN}}\hat\lambda)\,.
\label{susy10}
\eea
We can now use these standard ${\cal N}=1$ results in a generalised circle 
reduction to $d=9$.   We shall focus just on the pure supergravity 
multiplet in $d=9$, by performing a (consistent) truncation of the
 matter multiplet. The required reduction ansatz is obtained from
the arbitrary-dimension ansatz of appendix B by setting $m_1=m_2=m$
and $\phi_2=0=\chi$.  This gives
\bea
\hat\epsilon&=&e^{\fft12mz}e^{-\fft1{16\sqrt{14}}\phi_1}
\,\tilde\epsilon\,,\nn\\
\hat\lambda&=&\sqrt{\ft78}\,e^{-\fft12mz}
e^{\fft1{16\sqrt{14}}\phi_1}\tilde\lambda\,,\nn\\
\hat\psi_{\sst{10}}&=&-\,\ft{\sqrt7}8\,e^{-\fft12mz}
e^{\fft1{16\sqrt{14}}\phi_1}\td\gamma_{\sst{10}}\tilde\lambda\,,\nn\\
\hat\psi_a&=&e^{-\fft12mz}e^{\fft1{16\sqrt{14}}\phi_1}
{\Big(}\tilde\psi_a+\ft1{8\sqrt7}\,
\tilde\gamma_a\tilde\lambda{\Big)}\,,\nn\\
\hat\phi&=&\ft{\sqrt{14}}4\,\phi_1+4mz\,.\label{susyred}
\eea
The tildes signify that the fermions and the Dirac matrices 
are still ten-dimensional.  These can be related to the nine-dimensional 
quantities as follows:
\bea
&&\td\gamma_a=\gamma_a\times\sigma_{\sst 1}\,,\qquad 
\td\gamma_{\sst{10}}=\oneone\times\sigma_{\sst 2} \qquad\mbox{and} \qquad 
\hat\gamma_{\sst{11}}=\oneone\times\sigma_{\sst 3}\,,\nn\\
&&
\td\epsilon=\epsilon\times\eta\,, \qquad \td\lambda=
\lambda\times\sigma_{\sst 1}
\eta \qquad\mbox{and}\qquad \td\psi_a=\psi_a\times\eta\,,
\eea
where $\eta$ is a 2-component constant spinor. The chiral projections
(\ref{chiral}) imply that we must have $\sigma_{\sst 3}\eta=\eta$.  In
the following subsections, we present the resulting nine-dimensional
transformation rules in the Einstein frame and the string frame.

\subsubsection{$D=9$ supersymmetry in the Einstein frame}

   Reducing the ${\cal N}=1$, $D=10$ transformation rules, 
and setting $G_\2=\cF_\2=\frac1{\sqrt2}\,F_\2$, we obtain the
following nine-dimensional supersymmetry transformation rules:
\bea
\delta\lambda&=&-\,\ft1{2\sqrt2}\gamma^{\,\mu}\epsilon
\,\del_\mu\phi+\ft1{12\sqrt7}e^{\sqrt{\fft27}\phi}H_{\mu\nu\sigma}
\gamma^{\,\mu\nu\sigma}\epsilon+\ft{\rm i}{4\sqrt{14}}
e^{\frac1{\sqrt{14}}\phi}F_{\mu\nu}\gamma^{\,\mu\nu}\epsilon\nn\\
&&+\,\ft{4}{\sqrt7}m\,{\Big(}\ft1{\sqrt2}\gamma^{\,\mu}A_\mu
-{\rm i}e^{-\frac1{\sqrt{14}}\phi}{\Big)}\epsilon\,,\nn\\
\delta\psi_{\mu}&=&\nabla_\mu\epsilon-\ft1{84}
e^{\sqrt{\frac27}\phi}H_{\nu\sigma\rho}(\gamma_{\mu}^{\;\;\nu\sigma
\rho}-\ft{15}2\delta^\nu_\mu\gamma^{\sigma\rho})\epsilon
-\ft{\rm i}{28\sqrt2}e^{\frac1{\sqrt{14}}\phi}F_{\nu\sigma}
(\gamma_\mu^{\;\;\nu\sigma}-12\delta^\nu_\mu
\gamma^\sigma)\epsilon\nn\\
&&-\,\ft{4}{7\sqrt2}mA_\nu\gamma_\mu\gamma^\nu\epsilon
+\ft{4\rm i}7me^{-\frac1{\sqrt{14}}\phi}\gamma_\mu\epsilon\,,\nn\\
\delta e_\mu^{\;\;a}&=&\bar\epsilon\gamma^a\psi_\mu\,, \qquad
\delta\phi=-\,\sqrt2\,\bar\epsilon\,\lambda\,,\nn\\
\delta A_\mu&=&{\rm i}\sqrt2 e^{-\fft1{\sqrt{14}}\phi}\bar\epsilon(\psi_\mu
+\ft1{\sqrt7}\gamma_\mu\lambda)\,,\nn\\
\delta B_{\mu\nu}&=&-e^{-\sqrt{\fft27}\phi}\bar\epsilon
(2\gamma_{[\mu}\psi_{\nu]}
+\ft2{\sqrt7}\gamma_{\mu\nu}\lambda) - A_{[\mu}\delta A_{\nu]}\,,\label{susy9}
\eea
where we have dropped the ``1'' subscript on the scalar field. The field 
strengths
are $H_{\mu\nu\rho}=3\del_{[\mu}B_{\nu\rho]}-\ft32A_{[\mu}F_{\nu\rho]}$
and $F_{\mu\nu}=2\del_{[\mu}A_{\nu]}$\,.  This theory is an Abelian
gauged version of ${\cal N}=1, D=9$ supergravity.  We shall show that it
admits a supersymmetric $(\mbox{Minkowski})_6\times S^3$ vacuum solution.
We shall also obtain a time-dependent supersymmetric cosmological solution
in this theory.

\subsubsection{$D=9$ supersymmetry in the string frame}

The above transformation rules for the fermions are readily expressed
in terms of the fields of the string frame, using the formulae given in
appendix C. Specialised to nine dimensions, these are
\bea
&&g_{\mu\nu}=e^{\sqrt{\frac27}\phi_1}\td g_{\mu\nu}\,,\qquad F_\2=\wtd F_\2\,,
\qquad H_\3=\wtd H_\3\,, \qquad d\Phi+\sqrt8mA_\1=\wtd M_\1\,,\nn\\
&&\phi_1=-\,\sqrt{\ft87}\,\Phi\,,\quad
\epsilon=e^{\frac1{2\sqrt{14}}\phi_1}\td\epsilon\,, \quad
\lambda=e^{-\frac1{2\sqrt{14}}\phi_1} \td\lambda\,, \quad
\psi_{\mu}=e^{\frac1{2\sqrt{14}}\phi_1}\td\psi_{\mu}\,,
\eea
The fermionic transformation rules in the string frame then take the form
\bea
\delta\td\lambda&=&{\Big(}\ft1{\sqrt7}\wtd M_\mu\tilde\gamma^{\,\mu}
+\ft1{12\sqrt7}\wtd H_{\mu\nu\sigma}\td\gamma^{\,\mu\nu\sigma}
+\ft{\rm i}{4\sqrt{14}}\wtd F_{\mu\nu}\td\gamma^{\,\mu\nu}
-\,\ft{4\rm i}{\sqrt7}m{\Big)}\td\epsilon\,,\nn\\
\delta\td\psi_{\mu}&=&{\Big(}\wtd\nabla_{\mu}
-\ft17\wtd M_\nu\td\gamma_\mu\td\gamma^\nu
-\ft1{84}\wtd H_{\nu\sigma\rho}(\td\gamma_{\mu}^{\;\;\nu\sigma\rho}
-\ft{15}2\delta^\nu_\mu\td\gamma^{\sigma\rho})\nn\\
&&-\,\ft{\rm i}{28\sqrt2}\wtd F_{\nu\sigma}(\td\gamma_{\mu}^{\;\;\nu\sigma}
-12\delta^\nu_\mu\td\gamma^\sigma)+\ft{4\rm i}7m
\td\gamma_\mu{\Big)}\td\epsilon\,.
\eea

\section{Supersymmetric $M_{d-3}\times S^3$ and $M_{d-2}\times S^2$ vacua}

   The generalised Kaluza-Klein reduction gives rise to gauged
supergravities that admit supersymmetric vacuum solutions of the
form Minkowski$\times$Sphere \cite{kerimo2}.  The nine-dimensional
theory admits just a (Minkowski)$_6\times S^3$ vacuum of this kind,
supported by the $H_\3$ flux.  The theories in lower dimensions
admit (Minkowski)$_{d-3}\times S^3$ vacua supported by $H_\3$, and
(Minkowski)$_{d-2}\times S^2$ vacua supported by a 2-form $F_\2$.
In this section, we shall show that these vacua are all supersymmetric.

   Consider first the (Minkowski)$_{d-3}\times S^3$ solution supported by 
the $H_\3$ field. This is given by
\bea
ds^2_d&=&dx^\mu\,dx^\nu\,\eta_{\mu\nu} + \fft4{m^2\,(d-1)^2}\,
d\Omega_3^2\,,\nn\\
H_\3&=&\fft8{m^2\,(d-1)^2}\, \Omega_\3\,,\qquad \phi=0\,.\label{sol1}
\eea
If we lift the solution back to $D$
dimensions using the generalised reduction ansatz, it becomes the 
near-horizon geometry of a $(D-5)$-brane
supported by the field $\hat H_\3$.  To see this, we start with the
$(D-5)$-brane in $D$ dimensions, given by
\bea
d\hat s^2_{\sst D} &=& H^{-\fft2{D-2}}\, dx^\mu\,dx^\nu\,\eta_{\mu\nu} +
H^{\fft{D-4}{D-2}}\, (dr^2 + r^2\, d\Omega_3^2)\,,\nn\\
\hat H_\3 &=& 2Q\,\Omega_\3\,,\qquad 
\hat\phi = -\ft12\hat a\,\log H\,,\qquad H=1 + Q/r^2\,.
\eea
In the near-horizon limit, the additive constant 1 in $H$ is dropped.
Making the coordinate transformation $r^2/Q=e^{(D-2)\,m\,z}$, and letting
$Q=4/((D-2)^2\,m^2)$, we obtain
\bea
d\hat s_{\sst D}^2 &=& e^{2mz}\, \Big(dx^\mu\,dx^\nu\,\eta_{\mu\nu} + dz^2 +
\fft4{m^2\,(D-2)^2}\, d\Omega_3^2\Big)\,,\nn\\
\hat H_\3 &=& \fft8{m^2\, (D-2)^2}\, \Omega_\3\,,\qquad
\hat\phi = \fft4{\hat a}\, m z\,,\label{sol12}
\eea
which fits the reduction ansatz precisely, giving rise to the
lower-dimensional solution (\ref{sol1}).  

    The supersymmetry of the (Minkowski)$_{d-3}\times S^3$ solution is
easily established.  Firstly, since its lift to $D=d+1$ dimensions
gives the near-horizon limit of the $(D-5)$-brane, as discussed above,
it is manifest that {\it qua} $D$-dimensional solution, it will
preserve one half of the $D$-dimensional supersymmetry.  This halving
of supersymmetry comes about from the usual projection condition for
supersymmetry of the $(D-5)$-brane, $\hat\epsilon = \hat\Gamma_*\,
\hat \epsilon$, where $\hat\Gamma_*$ is built from the product of
Dirac matrices in the world-volume of the $(D-5)$-brane.  As is well
known, for any of the BPS brane solutions with metric given by
\be
d\hat s^2 = e^{2A}\, dx^\mu\, dx_\mu + e^{2B}\, dy^m\, dy^m\,,
\ee
the Killing spinors are given by
\be
\hat\ep = e^{\fft12 A}\, \hat\ep_0\,,\qquad 
\hat \Gamma_*\, \hat\ep_0 =\hat\ep_0
\,,\label{epform1}
\ee
where $\hat\ep_0$ is a constant spinor.  We see from (\ref{sol12}) that
$A= m z$, and hence the Killing spinors in $D$ dimensions take the form
\be
\hat\ep = e^{\fft12 m z}\, \hat\ep_0\,.\label{epform2}
\ee
Since this $z$ dependence matches precisely the $z$ dependence for
$\hat\epsilon$ in the generalised reduction ansatz (\ref{susyred}), it
immediately follows that the (Minkowski)$_{d-3}\times S^3$ solution
will be supersymmetric {\it qua} solution of the $d$-dimensional
gauged supergravity.

   Another class of supersymmetric vacuum is of the form (Minkowski)$_{d-2}
\times S^2$, supported by one of the two-form field strengths $F_\2^a$.
It is given by
\bea
ds^2_d &=& dx^\mu\,dx^\nu\,\eta_{\mu\nu} + \fft1{m^2\,(d-1)^2}\,
d\Omega_2^2\,,\nn\\
F_\2 &=& \fft{\sqrt2}{m\,(d-1)}\,\Omega_\2\,,\qquad
\phi=0\,.\label{sol2}
\eea
Lifting this solution back to $D$ dimensions, it becomes the
near-horizon limit of the $(D-4)$-brane supported by one of the field
strengths $\hat F_\2^a$.  The $(D-4)$-brane solution is given by
\bea
d\hat s^2_{\sst D} &=& H^{-\fft2{D-2}}\, dx^\mu\,dx^\nu\,\eta_{\mu\nu} +
H^{\fft{2(D-3)}{D-2}}\, (dr^2 + r^2\, d\Omega_2^2)\,,\nn\\
\hat F_\2 &=& \sqrt2\,Q\,\Omega_\2,\qquad
\hat \phi = -\ft12\hat a\,\log H \,,\qquad H=1 + Q/r\,.
\eea
In the near-horizon limit, the constant 1 in $H$ is
dropped. Making the coordinate transformation $r/Q=e^{(D-2)\,mz}$ and
setting $Q=1/(m\,(D-2))$ we have
\bea
d\hat s^2_{\sst D} &=&
e^{2m z}\, \Big( dx^\mu\,dx^\nu\,\eta_{\mu\nu} + dz^2 +
\fft1{m^2\, (D-2)^2}\, d\Omega_2^2\Big)\,,\nn\\
\hat F_\2 &=& \fft{\sqrt2}{m\,(D-2)}\,\Omega_\2\,,\qquad
\hat \phi = \fft4{\hat a}\,mz\,.\label{sol21}
\eea
This clearly fits the reduction ansatz exactly to give rise to
(\ref{sol2}). 

    Again, the supersymmetry of the solution as a lifted
$D$-dimensional configuration is manifest, since it is just the
near-horizon limit of a BPS $(D-4)$-brane. Its supersymmetry as a
solution in the $d=D-1$ dimensional gauged supergravity itself is
again easily seen, from the general form (\ref{epform1}) of the Killing
spinors in the lifted $(D-4)$-brane.  Thus we again find that the
$D$-dimensional Killing spinors are of the form (\ref{epform2}), and
so comparison with the generalised reduction ansatz (\ref{susyred})
for $\hat\epsilon$ shows that the (Minkowski)$_{d-2}\times S^2$
solution will be supersymmetric in the $d$-dimensional gauged
supergravity.

\section{Supersymmetric time-dependent solutions and pp-waves}
 
   In this section we construct a time-dependent solution of the new
gauged nine-dimensional supergravity, and we show that it is 
supersymmetric.  It can be thought of as a cosmological solution in
the gauged supergravity.

    The solution is of a form analogous to a standard domain
wall, except that here the ``transverse space coordinate'' is timelike
rather than spatial. It is easily seen that the configuration 
\bea
ds^2_9&=&-dt^2+(\ft87m\,t)^2dx^idx^i\,,\nn\\
e^{\fft1{\sqrt{14}}\phi}&=&\ft87m\,t\,. \label{dsol1}
\eea
solves the nine-dimensional equations of motion that follow from 
(\ref{ddlag1}).  Note that the form-fields are all zero in this solution.

   The fermionic transformation rules (\ref{susy9}) 
in this background reduce to 
\bea
\delta\lambda&=&-\ft1{2\sqrt2}\gamma^{\sst M}
(\del_{\sst M}\phi)\, \ep -\ft{4\rm i}{\sqrt7}m
e^{-\fft1{\sqrt{14}}\phi}\ep\,,\nn\\
\delta\psi_{\sst M}&=&\nabla_{\sst M}\, \ep 
+\ft{4\rm i}7me^{-\fft1{\sqrt{14}}\phi} \gamma_{\sst M}\, \ep\,,
\eea
and it is easily verified that (\ref{dsol1}) is supersymmetric.

    In the string frame, the metric in the solution (\ref{dsol1})
becomes simply the Minkowski metric
$ds^2_{\rm{str}}=\eta_{\sst{MN}}dx^{\sst M}dx^{\sst N}$, where
\be
t = \exp(\ft87 m\, x^0)\,.
\ee
The dilaton is a linear function of the redefined time; $\Phi = 
-4m x^0+$ constant.  

   The solution (\ref{dsol1}) is straightforwardly lifted to ten
dimensions, where it gives
\bea
ds^2_{\sst{10}}&=&e^{2mz}{\Big[}-(\ft87m\,t)^{-1/4}dt^2+(\ft87m\,t)^{7/4}
(dz^2+dx^idx^i){\Big]}\,,\nn\\
e^{\hat\phi}&=&e^{4mz}(\ft87m\,t)^{7/2}\,. \label{dsol2}
\eea
This can again be viewed as a time-dependent supersymmetric
cosmological solution, driven purely by the dilaton.  In the string
frame the metric is again Minkowskian, but now the dilaton is
linearly proportional to the light-cone coordinate $x^+$:
\be
ds_{\rm str}^2 = 2 dx^+\, dx^- + dx^i\, dx^i\,,\qquad
\Phi=x^+\,.
\ee
A metric-dilaton configuration of this kind was also discussed in
\cite{russo}.  It is straightforward to see that the solution
preserves half of the supersymmetry, with the Killing spinor given by
$\gamma_+\, \epsilon_0$ where $\epsilon_0$ is a constant
spinor.

     A further uplift to $D=11$ using the standard Kaluza-Klein formula
\be
ds^2_{\sst{11}}=e^{\fft16\hat\phi}ds^2_{\sst{10}}+e^{-\fft43\hat\phi}dy^2
\ee
yields the Ricci-flat solution
\be
ds^2_{\sst{11}}=-r^2dt^2+t^2dr^2+r^2t^2dx^idx^i+r^{-4}t^{-4}dy^2\,,
\label{dsol3}
\ee
where we have changed from the ten-dimensional coordinate $z$ to a 
new coordinate $r$ defined by
$r=e^{\fft43mz}(\ft87m\,t)^{1/6}$\,.  The metric (\ref{dsol3}) is a
pp-wave.  To see this, we
introduce new coordinates $X_+$ and $X_-$ defined by 
\be
r^2 \, t^2= X_+\,, \qquad  \frac{r}{t}=e^{2 X_-}\,,
\ee
in terms of which (\ref{dsol3}) becomes
\be
ds^2_{\sst{11}}=dX_+dX_-+X_+dx^idx^i+X^{-2}_+dy^2\,. \label{dsol5}
\ee
Thus, we conclude that in eleven dimensions the solution describes a
pp-wave.  

   The metric (\ref{dsol5}) is a particular example of a more 
general class of pp-waves, contained within the ansatz
\be
ds_{\sst D}=dX_+dX_-+X_+^{h_1}dx^{m_1}dx^{m_1}+X_+^{h_2}dy^{m_2}dy^{m_2}
+X_+^{h_3}dz^{m_3}dz^{m_3}+\cdots\,.\label{genmetan}
\ee
Here, we take the index ranges to be
\be
1\le m_1\, \le p_1\,,\qquad
p_1 +1\le m_2\, \le p_1+ p_2\,,\qquad \hbox{etc}.\,,
\ee
and so the total dimension is $D=2+p_1+p_2+\cdots$\,. 
The only non-vanishing vielbein components of the Riemann tensor 
for (\ref{genmetan}) are given by
\be
R_{m_i\, +\, m_j\, +}=-\ft12h_i(h_i-2)X_+^{-2}\delta_{m_i\,m_j}\,.
\ee
Thus (\ref{genmetan}) is Ricci-flat if
\be
0=\sum_{i=1}p_ih_i(h_i-2)\,. \label{rc}
\ee
The pp-wave (\ref{dsol5}) that resulted from lifting our
time-dependent cosmological solution to $D=11$ is the special case
with
\be
p_1=8\,, \qquad h_1=1\,,\qquad
p_2=1\,, \qquad h_2=-2\,,
\ee
which clearly satisfies (\ref{rc}).

\section{Conclusions}

         In this paper, we have obtained generalised Kaluza-Klein
reductions of the low-energy effective actions of string theories
involving the metric, the dilaton, a 3-form field strength and a 2-form
field strength.  The generalised reduction gauges two global symmetries,
namely the homogeneous scaling symmetry of the equations of motion,
and also the dilaton shift symmetry of the Lagrangian.  The resulting
dimensionally-reduced theory has a positive scalar potential, in
the form of a single-exponential of the lower-dimensional dilaton.
We showed that the reduction is supersymmetric, by explicitly deriving
the lower-dimensional supersymmetry transformation rules.

   Although it might seem somewhat perverse to perform generalised
reductions of the kind we have considered in this paper, they are
actually related by U-duality to more conventional reductions that
have been considered extensively in the past.  Specifically, a
generalised reduction involving the global shift symmetry of the axion
in the type IIB theory has been used in order to establish a T-duality
between the type IIB theory and the massive type IIA theory
\cite{brgpt}.  The S-duality of the type IIB theory implies that one
should also consider $SL(2,R)$-related generalised reductions
\cite{mo}, which will involve the global shift symmetry of the
dilaton.  When one extends the discussion of non-perturbative
dualities to lower dimensions, the underlying global Cremmer-Julia
type symmetries can only be interpreted as strictly internal
symmetries if one also makes use of the scaling symmetry of the
equations of motion that homogeneously scales the Lagrangian.  Thus it
is very natural to consider generalised reductions of the kind we have
studied in this paper.

       The new supergravities have the interesting feature that they
all admit supersymmetric vacuum solutions of the form
(Minkowski)$\times S^3$, and in some cases also (Minkowski)$\times
S^2$.  These solutions provide novel compatifications of higher
dimensional string theories.  Furthermore, owing to the positivity of
the scalar potential, the supergravities we have obtained admit
time-dependent cosmological solutions that preserve half of the
supersymmetry.  Lifting these solutions back to $D=10$, they yield
supersymmetric time-dependent solutions driven purely by the dilaton,
with no form-field fluxes.  Under a further lifting to eleven
dimensions, these time-dependent solutions become supersymmetric
pp-waves.  It would be interesting to study string theory and M-theory
in these simple but non-trivial backgrounds.

\begin{appendix}

\section{Bosonic reduction ansatz; Einstein frame}

   We begin by reducing the $D=d+1$ dimensional Ricci tensor to $d$
dimensions by using the metric ansatz in (\ref{genredans}).
We choose the natural vielbein basis
\be
{\hat{e}}^a=e^{m_2z+\alpha\varphi}e^a,\qquad
{\hat{e}}^z=e^{m_2z+\beta\varphi}
(dz+\cA_\1)\,.\label{ba1}
\ee
Thus we have
\be
\hat e_{\sst M}^{\;\;\sst {A}}=e^{m_2z}\left(
\begin{array}{cc} e^{\alpha\varphi} e_\mu^{\;\;a} &
e^{\beta\varphi}\cA_\mu \\
0 & e^{\beta\varphi} \end{array}\right)\,, \qquad
\hat e_{\sst A}^{\;\;\sst M}=e^{-m_2z}\left(
\begin{array}{cc} e^{-\alpha\varphi}e_a^{\;\;\mu} &
-e^{-\alpha\varphi}\cA_a \\
0 & e^{-\beta\varphi} \label{spin} \end{array}\right)
\,.\label{ve}
\ee
The determinant of the metric is
\be
\sqrt{-\hat g}=e^{(d+1)m_2z+(\beta+d\alpha)\varphi}\sqrt{-g}
=e^{(d+1)m_2z+2\alpha\varphi}\sqrt{-g}\,.
\label{det}
\ee

Using the first Cartan structure equation with zero torsion, $d\hat
e^{\sst A}=-\hat\omega^{\sst A}_{\;\;\sst B}\wedge\hat e^{\sst B}$, we obtain
the spin connections
\bea
{\hat{\omega}}^a_{\;\;b}&=&{\omega}^a_{\;\;b}+e^{-(m_2z+\alpha\varphi)}
{\Big(} (\alpha\del_b\varphi-m_2\cA_b)\,
\hat e^a-(\alpha\del^{\,a}\varphi-m_2\cA^a)\,\hat e_b{\Big)}\nn\\
&&-\ft12e^{-m_2z+(\beta-2\alpha)\varphi}\cF^a_{\;\;b}\,\hat e^z,
\label{co1}\\
{\hat{\omega}}^a_{\;\;z}&=&e^{-(m_2z+\alpha\varphi)}
(m_2\cA^a-\beta\del^{\,a}
\varphi)\,\hat e^z-\ft12e^{-m_2z+(\beta-2\alpha)\varphi}
\cF^a_{\;\;b} \,\hat e^b+m_2e^{-(m_2z +\beta\varphi)}
\hat e^a\,. \nn
\eea
From the curvature 2-forms 
$\hat\Theta^{\sst A}{}_{\sst B}=d\hat\omega^{\sst
A}{}_{\sst B}+\hat\omega^{\sst A}{}_{\sst C}\wedge\hat\omega^{\sst
C}{}_{\sst B} = \frac{1}{2}\hat R^{\sst A}{}_{\sst {BCD}}\hat e^{\sst
C}\wedge\hat e^{\sst D}$, we obtain the Ricci tensor with vielbein
components
\bea
\hat R_{ab}&=&e^{-2(m_2z+\alpha\varphi)}
{\Big(}R_{ab}-\ft12\del_a\varphi\,\del_b\varphi
-\alpha\eta_{ab}\,\square\varphi\nn\\
&&+ {\alpha}m_2(d-1)(\cA^c\del_{\,c}\varphi\,\eta_{ab}
-\cA_a\del_b \varphi-\cA_b\del_a\varphi) \nn \\
&& +\ft12m_2(d-1)(\na_a\cA_b+\na_b\cA_a)+
m_2\na_c\cA^c{\eta}_{ab}+m_2^2(d-1)
(\cA_a\cA_b-\cA^2_\1{\eta}_{ab}){\Big)} \nn \\
&&-\;m_2^2(d-1)e^{-2(m_2z+\beta\varphi)}{\eta}_{ab}
-\ft12e^{-2(m_2z+d\alpha\varphi)}\cF_a{}^c
\cF_{bc}\,, \nn\\
\hat R_{az}&=&e^{-2m_2z+(d-3)\alpha\varphi}{\Big(}
\ft12\nabla^b{\big(}e^{-2(d-1)\alpha\varphi} 
\cF_{ab}{\big)}+m_2(d-1)(\beta\del_a\varphi-m_2\cA_a)
{\Big)} \nn\\
&& -\,\ft12m_2(d-1)e^{-2m_2z-(d+1)\alpha\varphi}
\cA^b\cF_{ab}\,, \nn\\
\hat R_{zz}&=&e^{-2(m_2z+\alpha\varphi)}{\Big(}
-\beta\,\square\varphi+m_2\nabla_c\cA^c+ m_2\beta(d-1)
\cA^b\del_b\varphi-m_2^2(d-1)\cA^2_\1 {\Big)} \nn \\
&&+\,\ft14e^{-2(m_2z+d\alpha\varphi)}\cF_\2^2\,. 
\label{ricci}
\eea
The Ricci scalar is
\bea
\hat R&=&e^{-2(m_2z+\alpha\varphi)}{\Big(}
R-2\alpha\,\square\varphi-\ft12{(\del\varphi)}^2
+2m_2d\,\nabla_a\cA^a-m_2^2d\,(d-1)\cA^2_\1{\Big)}\nn \\
&& -\,e^{-2m_2z}{\Big(}m_2^2\,d\,(d-1)e^{-2\beta\varphi}+
\ft14e^{-2d\alpha\varphi} \cF^2_\2{\Big)}\,.
\label{rscalar}
\eea
The reduced Ricci components in (\ref{ricci}) have been simplified through
use of the relations (\ref{diag}).

   The Laplacian operator acting on the $D$-dimensional dilaton is given by
\be
e^{2m_2z+2\alpha\varphi}\,\widehat{\square}\hat\phi
=\square\phi -m_2(d-1){\Big (}\cA^\mu\del_\mu\phi-
\frac4{\hat a}\,m_1\,(\cA^2_\1+e^{2(d-1)\alpha\varphi}){\Big )}
-\,\frac4{\hat a}\,m_1\nabla_\mu\cA^\mu\,,\label{box}
\ee
where $\hat\phi=\phi+\frac4{\hat a}\,m_1z$, as given by
(\ref{genredans}).

   The vielbein components of the various $D$-dimensional 
antisymmetric tensors reduce according to
\bea 
\hat H_{a_1{\cdots}a_n}&=&e^{-(m_2 + (n-1)m_1)z-n\alpha\varphi}\,
H_{a_1{\cdots}a_n}\,,\nn\\ \hat H_{a_1{\cdots}a_{n-1}z}&=&
e^{-(m_2 + (n-1)m_1)z+(d-n-1)\alpha\varphi}\,
H_{a_1{\cdots}a_{n-1}}\,.\label{comp}
\eea

\section{Fermionic reduction ansatz in $D\le10$; Einstein frame}

   In this appendix we provide an arbitrary dimensional generalised
ansatz that reduces the fermions in $D=d+1$ to $d$ dimensions.  The
generalised ansatz we are constructing is such that the standard $S^1$
reduction $(m_1=0=m_2)$ reduces canonical fermionic kinetic terms with a
normalization as
\be
\hat e^{-1}\hat{\cal L}=\kappa(\hat{\bar\Psi}_{\sst M}
\hat\gamma^{\sst{MNP}}
\widehat\nabla_{\sst N}\hat\Psi_{\sst P}+
\hat{\bar\lambda}\hat\gamma^{\sst M}
\widehat\nabla_{\sst M}\hat\lambda)
\ee
to canonical kinetic terms
\be
e^{-1}{\cal L}=\kappa(\bar\Psi_{\mu}\gamma^{\mu\nu\rho}
\nabla_{\nu}\Psi_{\rho}+\bar\lambda\gamma^{\mu}\nabla_{\mu}\lambda
+\bar\chi\gamma^{\mu}\nabla_{\mu}\chi)+\mbox{rest}\,.
\ee
Here $\kappa$ is an arbitrary coefficient.  Performing the split of
the gravitino as $\hat\psi_{\sst A}=(\hat\psi_a,\hat\psi_{\sst D})$ an
ansatz that accomplishes this is
\bea
\hat\epsilon&=&e^{\frac12m_2z}e^{\frac12\alpha\varphi}\,\epsilon\,,\nn\\
\hat\lambda&=&\ft1{\sqrt{D-2}}\,e^{-\frac12m_2z}e^{-\frac12\alpha\varphi}
(\chi+\sqrt{D-3}\,\lambda)\,,\nn\\
\hat\psi_{\sst D}&=&\ft{\sqrt{D-3}}{D-2}e^{-\frac12m_2z}e^{-\frac12
\alpha\varphi}\gamma_{\sst D}(\sqrt{D-3}\,\chi-\lambda)\,,\nn\\
\hat\psi_a&=&e^{-\frac12m_2z}e^{-\frac{1}{2}\alpha\varphi}(\psi_a
-\ft1{(D-2)\sqrt{D-3}}\gamma_a(\sqrt{D-3}\,\chi-\lambda))\,,\nn\\
\hat\phi&=&\sqrt{\ft{D-3}{D-2}}\,\phi_1+\ft1{\sqrt{D-2}}\phi_2
+\sqrt{2(D-2)}\,m_1z\,,\nn\\
\varphi&=&-\,\ft1{\sqrt{D-2}}\phi_1+\sqrt{\ft{D-3}{D-2}}\phi_2\,.
\eea
Note that, here and elsewhere in this paper our convention is always
$\alpha>0$\,.  A consistent truncation of the matter multiplet can be
obtained by setting $m_1=m_2$ and $\phi_2=0=\chi$\,.

\section{Einstein-frame to string-frame conversion}

    The $D$-dimensional Lagrangian in the Einstein frame is given by
\bea
e^{-1}{\cal L}&=&R-\ft12(\del\phi)^2-\ft1{12}e^{\hat a\phi}
H_\3^{2}-\ft14e^{\frac12\hat a\phi}(F_\2^{a})^2
-\ft12\bar\Psi_{\sst M}\gamma^{\sst{MNP}}\nabla_{\sst N}\Psi_{\sst P}
-\ft12\bar\lambda\,\gamma^{\sst M}\nabla_{\sst M}\lambda\nn\\
&&-\ft1{2\sqrt{2}}\bar\lambda\gamma^{\sst{N}}\gamma^{\sst{M}}
\Psi_{\sst{N}}\partial_{\sst{M}}\phi
+\cdots\,,
\eea
where $\hat a=\sqrt{\fft8{D-2}}$, and where we have omitted additional
interaction and four-fermi terms.  This may be mapped to the string
frame Lagrangian
\bea
\td e^{-1}\wtd {\cal L}&=&e^{-2\Phi}{\Big(}\wtd R+4(\del\Phi)^2-\ft1{12}
\wtd H_\3^{2}-\ft14(\wtd F_\2^{a})^2
-\ft12\bar{\td\Psi}_{\sst M}\wtd\gamma^{\sst{MNP}}\wtd\nabla_{\sst N}
\td\Psi_{\sst P}-\ft12\bar{\td\lambda}\,\wtd\gamma^{\sst M}
\wtd\nabla_{\sst M}\td\lambda\nn\\
&&-(\bar{\td\Psi}_{\sst{N}}\wtd\gamma^{\sst{N}}\td\Psi^{\sst{M}}
-\fft{\hat a}{2\sqrt{2}}\bar{\td\lambda}\wtd\gamma^{\sst{N}}
\wtd\gamma^{\sst{M}}\td\Psi_{\sst{N}})\partial_{\sst{M}}\Phi+\cdots{\Big)}\,,
\eea
by the transformations
\bea
g_{\sst{MN}}&=&e^{\fft12 \hat a\phi}\,\td g_{\sst{MN}}\,,\qquad
H_{\sst{MNP}}=\wtd H_{\sst{MNP}}\,,\qquad
F^a_{\sst{MN}}=\wtd F^a_{\sst{MN}}\,,\qquad
\phi = -\hat a\, \Phi\,,\nn\\
\epsilon&=&e^{\fft18 \hat a\phi}\,\td\epsilon\,,\qquad
\lambda=e^{-\fft18 \hat a \phi}\td\lambda\,,\qquad
\Psi_{\sst M}=e^{\fft18 \hat a \phi}\td\Psi_{\sst M}\,.
\eea
Note that $\gamma_{\sst M}
=e^{\frac14\hat a\phi}\,\td\gamma_{\sst M}$ i.e. $\gamma_{\sst A}
=\td\gamma_{\sst A}$\,.  Furthermore, we have made use of
the $D$-dimensional
Majorana flip properties $\bar\psi\gamma^{\sst{M}}\chi=-\bar\chi
\gamma^{\sst{M}}\psi$ and $\bar\psi\gamma^{\sst{MNP}}\chi=\bar\chi
\gamma^{\sst{MNP}}\psi$ for any two anti-commuting spinors $\psi$ and
$\chi$.

   The bosonic reduction ans\"atze in the string frame are considerably
simpler than their Einstein-frame counterparts.  The reduction of the
$D=d+1$ dimensional Ricci tensor is given by
\bea
\hat R_{ab}&=&R_{ab}+\ft1{\sqrt2}\nabla_a\del_b\varphi-\ft12
\del_a\varphi\del_b\varphi-\ft12e^{-\sqrt2
\varphi}{\cal F}_{ac}\,{\cal F}_b^{\;\,c}\,,\nn\\
\hat R_{az}&=&\ft12e^{\sqrt2\varphi}\,\nabla^b
(e^{-\frac3{\sqrt2}\varphi}{\cal F}_{ab})\,,\nn\\
\hat R_{zz}&=&\ft1{\sqrt2}\,\square\varphi-\ft12
(\del\varphi)^2+\ft14e^{-\sqrt2\varphi}{\cal F}^{\,2}_{(2)}\,,\nn\\
\hat R&=&R+\sqrt2\,\square\varphi-(\del\varphi)^2
-\ft14e^{-\sqrt2\varphi}{\cal F}^{\,2}_{(2)}\,.
\eea

   Some useful formulae for the reduction of the scalar fields are:
\bea
\widehat{\square}\hat\Phi=\widehat{\square}{\Big(}\Phi-\frac{\varphi}
{\sqrt8}-\ft12(D-2)mz{\Big)}&=&\square\Phi-\ft1{\sqrt8}
\,\square\varphi-\ft1{\sqrt2}(\del_{\mu}\varphi\,\del^{\,\mu}
\Phi-\ft1{\sqrt8}(\del\varphi)^2)\nn\\
&&-\,\ft12m(d-1)(\ft1{\sqrt2}\,{\cal A}^{\mu}\del_{\mu}
\varphi-\nabla_{\mu}{\cal A}^{\mu})\,,
\eea
\bea
(\del\hat\Phi)^2&=&(\del\Phi)^2+\ft18(\del\varphi)^2
-\ft1{\sqrt2}\,\del_{\mu}\varphi\,\del^{\,\mu}\Phi+m(d-1)
{\cal A}^{\mu}(\del_{\mu}\Phi-\ft1{\sqrt8}\del_{\mu}\varphi)\nn\\
&&+\,\ft14m^2(d-1)^2({\cal A}_\1^2+e^{\sqrt2\varphi})\,,
\eea
\bea
\hat e_a^{\;\,\sst M}\hat e_b^{\;\,\sst N}\widehat\nabla_{\sst M}
\del_{\sst N}\hat\Phi&=&\nabla_a\del_b\Phi-\ft1{\sqrt8}\nabla_a
\del_b\varphi+\ft14m(d-1)(\nabla_a{\cal A}_b+\nabla_b{\cal A}_a)\,,
\nn\\
\hat e_a^{\;\,\sst M}\hat e_z^{\;\,\sst N}\widehat\nabla_{\sst M}
\del_{\sst N}\hat\Phi&=&-\,\ft12e^{-\frac1{\sqrt2}\varphi}
{\cal F}_a^{\;\,b}(\del_b\Phi-\ft1{\sqrt8}\,\del_b\varphi)
-\ft1{2\sqrt2}m(d-1)e^{\frac1{\sqrt2}\varphi}\del_a\varphi\nn\\
&&-\,\ft14m(d-1)e^{-\frac1{\sqrt2}\varphi}{\cal F}_{ab}{\cal A}^b\,,
\nn\\
\hat e_z^{\;\,\sst M}\hat e_z^{\;\,\sst N}\widehat\nabla_{\sst M}
\del_{\sst N}\hat\Phi&=&-\,\ft1{\sqrt2}\del^{\,\mu}\varphi
\,(\del_{\mu}\Phi-\ft1{\sqrt8}\,\del_{\mu}\varphi)
-\ft1{2\sqrt2}m(d-1){\cal A}^{\mu}\del_{\mu}\varphi\,.
\eea

\end{appendix}


\begin{thebibliography}{99}


\bm{szmodel}  A. Salam and E. Sezgin,
{\it Chiral compactification on Minkowski$\times S^2$ Of $N=2$ 
Einstein-Maxwell supergravity in six dimensions,}
Phys.\ Lett.\ {\bf B147}, 47 (1984).

\bm{abpq1} Y. Aghababaie, C.P. Burgess, S.L. Parameswaran and F. Quevedo,
{\it SUSY breaking and moduli stabilization from fluxes in gauged $6D$ 
supergravity,} JHEP {\bf 0303}, 032 (2003), hep-th/0212091.

\bm{abpq2} Y. Aghababaie, C.P. Burgess, S.L. Parameswaran and F. Quevedo,
{\it Towards a naturally small cosmological constant from branes 
in $6D$ supergravity,}
Nucl.\ Phys.\ {\bf B680}, 389 (2004),  hep-th/0304256.

\bm{cgp} M. Cveti\v c, G.W. Gibbons and C.N. Pope,
{\it A string and M-theory origin for the Salam-Sezgin model,}
Nucl.\ Phys.\ {\bf B677}, 164 (2004), hep-th/0308026.

\bm{gp} G.W. Gibbons and C.N. Pope,
{\it Consistent $S^2$ Pauli reduction of six-dimensional
chiral gauged Einstein-Maxwell supergravity,} hep-th/0307052.

\bm{lav} I.V. Lavrinenko, H. L\"u and C.N. Pope,
{\it Fibre bundles and generalised dimensional reductions,}
Class.\ Quant.\ Grav.\  {\bf 15}, 2239 (1998), hep-th/9710243.

\bibitem{kerimo1} J. Kerimo and H. L\"u,
{\it New $D=6$, ${\cal N}=(1,1)$ gauged supergravity with supersymmetric
(Minkowski)$_4\times S^2$ vacuum},
Phys.\ Lett.\ {\bf B576}, 219 (2003), hep-th/0307222.

\bibitem{kerimo2} J. Kerimo, J.T. Liu, H. L\"u and C.N. Pope,
{\it Variant ${\cal N}=(1,1)$ Supergravity and (Minkowski)$_4\times S^2$
Vacua}, hep-th/0401001.

\bibitem{brgpt}
E. Bergshoeff, M. de Roo, M.B. Green, G. Papadopoulos and P.K. Townsend,
{\it Duality of Type II 7-branes and 8-branes,}
Nucl.\ Phys.\ {\bf B470}, 113 (1996), hep-th/9601150.

\bibitem{clpst}
P.M. Cowdall, H. L\"u, C.N. Pope, K.S. Stelle and P.K. Townsend,
{\it Domain walls in massive supergravities,} Nucl.\ Phys.\ {\bf B486},
49 (1997), hep-th/9608173.

\bibitem{mo} P. Meessen and T. Ortin,
{\it An $Sl(2,Z)$ multiplet of nine-dimensional type
II supergravity theories}, Nucl.\ Phys.\ {\bf B541}, 195 (1999),
hep-th/9806120.

\bm{berg} E. Bergshoeff, T. de Wit, U. Gran, R. Linares and D. Roest,
{\it (Non-)Abelian gauged supergravities in nine dimensions,}
JHEP {\bf 0210}, 061 (2002), hep-th/ 0209205.

\bibitem{hlw}
P.S. Howe, N.D. Lambert and P.C. West,
{\it A new massive type IIA supergravity from compactification,}
Phys.\ Lett.\ {\bf B416}, 303 (1998), hep-th/9707139.

\bm{romans10} L.J. Romans,
{\it Massive $N=2a$ supergravity in ten-dimensions,}
Phys. Lett. {\bf B169}, 374 (1986).


\bibitem{russo} G. Papadopoulos, J.G. Russo and A.A. Tseytlin,
{\it Solvable model of strings in a time-dependent plane-wave background,}
Class.\ Quant.\ Grav.\  {\bf 20}, 969 (2003), hep-th/0211289.


\end{thebibliography}
\end{document}